\documentclass[preprint2]{aastex}

\slugcomment{}

\shorttitle{Angle-Dependent SSC Model}
\shortauthors{Jamil and B\"ottcher}

\def\bfield{{\textbf{B}-field}}

\begin{document}


\title{An Angle Dependent SSC Model for Relativistic Jet Sources}


\author{O. Jamil\altaffilmark{1} and M. B\"ottcher\altaffilmark{1}}
\affil{Astrophysical Institute, Department of Physics and Astronomy, 
Ohio University, Athens, OH, USA}


\altaffiltext{1}{Astrophysical Institute, Department of Physics and Astronomy, 
Ohio University, Athens, OH, USA}


\begin{abstract}
We report on the development of a numerical code to calculate the 
angle-dependent synchrotron + synchrotron self-Compton radiation 
from relativistic jet sources with partially ordered magnetic fields
and anisotropic particle distributions. Using a multi-zone radiation
transfer approach, we can simulate magnetic-field configurations 
ranging from perfectly ordered (unidirectional) to randomly oriented 
(tangled). We demonstrate that synchrotron self-Compton model fits
to the spectral energy distributions (SEDs) of extragalactic jet sources
may be possible with a wide range of magnetic-field values, depending
on their orientation with respect to the jet axis and the observer.
This is illustrated with the example of a spectral fit to the SED
of Mrk~421 from multiwavelength observations in 2006, where acceptable
fits are possible with magnetic-field values varying within a range of 
an order of magnitude for different degrees of B-field alignment and 
orientation. 
\end{abstract}





\section{Introduction}
\label{sec:intro}
Blazars form one of the most energetically extreme classes of
Active galactic nuclei (AGN). Blazars can be observed in all 
wavelengths, ranging from radio all the way up to $\gamma$-rays. 
Their spectral energy distribution (SED) is characterized by two
broad non-thermal components, one from radio through optical,
UV, or even X-rays, and a high-energy component from X-rays
to $\gamma$-rays. In addition to spanning across all observable 
frequencies, blazars are also highly variable across the electromagnetic
spectrum, with timescales ranging down to just a few minutes at 
the highest energies. 

There are two fundamentally different approaches to model the
SEDs and variability of blazars, generally referred to as 
leptonic and hadronic models \citep[see, e.g.,][for a review
of blazar models]{boettcher07}. In the case of leptonic
models, where leptons are the primary source of radiation,
synchrotron, synchrotron self-Compton (SSC), and external-Compton (EC) 
radiation mechanisms are employed to explain the blazar SED \citep[see, 
e.g.,][]{marscher85,maraschi92,dermer92,ghisellini96}. The focus of the
present study is also on a leptonic model. In hadronic models, the
low-energy SED component is still produced by synchrotron emission 
from relativistic electrons, while the high-energy component is
dominated by the radiative output from ultrarelativistic protons,
through photo-pion induced cascades and proton synchrotron emission
\citep[e.g.,][]{mannheim92,muecke01,muecke03}. One aspect common 
to all blazar models is a relativistic jet oriented at a small angle
with respect to our line of sight, resulting in relativistic Doppler
boosting and the shortening of observed variability time scales.

Given computational limitations, the complex physical processes in 
relativistic jets can, realistically, only be evaluated with certain
simplifying approximations. In order to facilitate analytical as well 
as numerical calculations, the two most common approximations employed
in blazar jet models are to assume that the magnetic (\textbf{B}) field 
is randomly oriented and tangled, and that the lepton momentum
distribution is isotropic in the comoving frame of the high-energy
emission region. These two assumptions greatly simplify the evaluation 
of the synchrotron and Compton emission by eliminating various 
integrals over the interaction and scattering angles. However,
there is increasing evidence \citep{attridge99,lyutikov05,marscher08}
for a fairly well defined helical \textbf{B}-field structure within
AGN jets. These observations also suggest a spine-sheath geometry
for AGN jets. The differential velocity profiles within the
jet is expected to create anisotropies in the particle distributions. 
It is therefore important to explore jet models where we can not only 
simulate an ordered \textbf{B}-field, but also study the resulting 
radiation behaviour with anisotropic lepton distributions.

\subsection{B-Field estimates}
\label{subsec:bestimate}
The standard approach to diagnosing the magnetic field properties 
is via synchrotron polarization. If the underlying distribution of emitting 
electrons is a power-law with power-law index $p$, the maximum degree of 
synchrotron polarization is given by:
\begin{eqnarray}
  \label{eq:synchpol}
  \Pi = \frac{P_{\perp}(\nu)-P_{\parallel}(\nu)}{P_{\perp}(\nu)+P_{\parallel}(\nu)}
= \frac{p+1}{p+\frac{7}{3}} 
\end{eqnarray}
where $P_{\perp}(\nu)$ and $P_{\parallel}(\nu)$ are the synchrotron 
power per unit frequency in directions perpendicular and parallel to the 
projection of the magnetic field on the plane of the sky. Using equation 
\ref{eq:synchpol} we can see that for a power-law index of $p = 3$, the 
degree of polarization can be as high as 75\%.

It is therefore possible to estimate the magnetic field orientation
based on polarization measurements, but an estimate of the field strength
usually requires the consideration of flux and spectral properties of the
synchrotron emission. Furthermore, polarization measurements are notoriously
difficult (and even barely feasible at frequencies higher than optical), and
may often not give realistic results due to Faraday rotation and depolarization
along the line of sight. In this work, we are interested in taking a complementary
approach to estimating the magnetic field orientation where the difference in 
observed flux levels of the spectrum can give an estimate of how the magnetic 
orientation may be changing.

\begin{figure}[h]
  \begin{center}
    \includegraphics[width=0.45\textwidth]{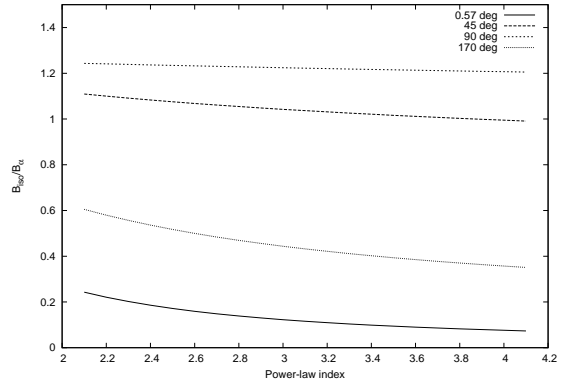}
    \caption{The $B_{iso}/B_{\alpha}$ ratio shows how the B-field
      estimate can vary depending on the electron power-law
      index and whether one assumes an
      isotropic B-field or a specific pitch angle.}
    \label{fig:BEstimate}
  \end{center}
\end{figure}

The principle behind this approach can be demonstrated when one compares the
\textbf{B}-field estimates based on a power-law distribution of
electrons with an arbitrary power-law index, $p$ , and pitch angle, $\alpha$. The comparison of synchrotron emission coefficients for a power-law
distribution of electrons with and without
pitch-angle ($\alpha$) dependence gives us a measure of how the estimated magnetic field strength can differ. The emission coefficients can be found in \cite{longair94} and are
given by (in the units of $\mathrm{erg~ s^{-1} cm^{-3}Hz^{-1}}$):
\begin{eqnarray}
j_{\alpha}(\nu)&=&\frac{\sqrt{3}\mathrm{e}^3\mathrm{B}\kappa\sin\alpha}
       {2 m_e c^2(p+1)}\left(\frac{4\pi\nu
         m_ec}{3e\mathrm{B}\sin\alpha} \right)^{-(p-1)/2}
       \nonumber \\
      &\times &\Gamma(\frac{p}{4}+\frac{19}{12})\Gamma(\frac{p}{4}-\frac{1}{12})  
\label{eqn:synchJalpha}  
\end{eqnarray}
and
\begin{eqnarray}
j_{\mathrm{iso}}(\nu) &=& \frac{\sqrt{3}\mathrm{e}^3\mathrm{B}\kappa}
       {4 m_e c^2(p+1)}\left(\frac{4\pi\nu
         m_ec}{3e\mathrm{B}} \right)^{-(p-1)/2} \nonumber \\
&\times& \frac{\Gamma(\frac{p}{4}+\frac{19}{12})\Gamma(\frac{p}{4}-\frac{1}{12})
         \Gamma(\frac{p}{4}+\frac{5}{4})}{\Gamma(\frac{p}{4}+\frac{7}{4})}
       ~ . 
\end{eqnarray}
Where $\kappa$ is the electron distribution power-law normalization. The above two 
expressions can be solved for the magnetic field to obtain:
\begin{eqnarray}
  \label{eqn:bestimate}
  \frac{B_{\mathrm{iso}}}{B_{\alpha}} = \left(
  \frac{2}{\sqrt{\pi}}\frac{(\sin\alpha)^{\frac{p-1}{2}}\Gamma(\frac{p}{4}+\frac{7}{4})}{
    \Gamma(\frac{p}{4}+\frac{5}{4})}\right)^{2/(p+1)}
  \ .
\end{eqnarray}
This gives an estimate of how, for a given luminosity, the \textbf{B}-field estimates
can differ depending upon whether we assume an isotropic pitch angle approximation or 
a given pitch angle 
(which, in the case of relativistic electrons, is equal to the angle
between the magnetic field and the line of sight). 
The above 
relation is only applicable in the optically thin regime. We can see in figure
\ref{fig:BEstimate} that depending on the pitch angle assumption, and
the electron distribution power-law index, the $B_{iso}/B_{\alpha}$ fraction can
range from 0 
(there is negligible synchrotron emission along an ordered magnetic field)
to $\sim 1.25$. Because the Compton emissivity is approximately 
isotropic for an isotropic distribution of electrons, the ratio $F_{SSC}/F_{Sy}$ 
will change with the pitch angle. It is therefore important to see how the overall 
synchrotron and synchrotron-self Compton spectra differ with well ordered magnetic 
fields. One point worth noting is that in our set-up the lower limit on \bfield\ 
orientation is limited by the $\delta(\Omega_{sy}-\Omega_e)$ approximation (see 
section \ref{sec:synch}); a magnetic field perfectly aligned with the observing 
direction will give zero output. However, in a more rigorous treatment, the lower 
limit on the minimum angle, $\alpha$, for the \bfield\ orientation will be determined 
by 
the relativistic beaming characteristic of synchrotron emission along
an electron's direction of motion into a cone of opening angle $\alpha_{\rm sy} \sim
1/\gamma$, 
where $\gamma$ is the electron Lorentz factor. In the 
case of optical frequencies 
and magnetic fields of $B \sim 1$~G, 
$\alpha_{min}$ will be the order of $10^{-4}$, while 
in the X-ray regime $\alpha_{min} \sim 10^{-5}$. Using the relation in equation 
\ref{eqn:bestimate} it is possible to estimate the effects on the \bfield\ estimates 
between this value of $\alpha_{min}$ and an isotropic magnetic field. For a power-law 
index of 3, $\alpha_{min}$ gives $B_{iso}/B_{\alpha}$ value of 0.0039. 

\section[]{The Model}
\label{sec:desc}
The following section briefly outlines our model, including the
synchrotron radiation and Compton scattering treatments followed as
well as the numerical techniques used to implement them.

\subsection[]{Volume, B-field, and Distributions}
\label{sec:volume}

\begin{figure}[h]
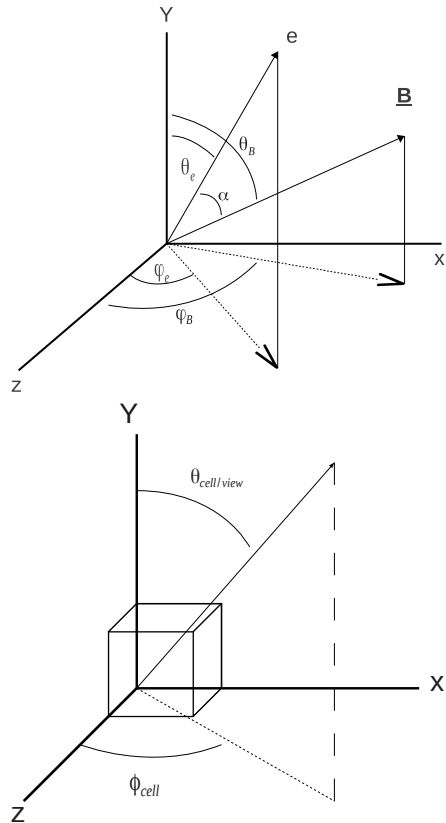

\begin{center}
  $\begin{array}{c}
   \includegraphics[width=0.35\textwidth]{figure2.epsi} \\ \includegraphics[width=0.35\textwidth]{figure3.epsi}\\    
  \end{array}$
\end{center}
  \caption{Top: The electron and \bfield\ interaction angles. Bottom: Cell/Jet geometry 
illustrating how the photon spectrum can be observed as a function of various angles, 
$\theta_{cell}$ and $\phi_{cell}$, with respect to the cell. The electron and photon 
distributions are defined with respect to the cell.} 
\label{fig:eleB}
\end{figure}

\begin{figure}[h]
\begin{center}
  \includegraphics[width=0.4\textwidth]{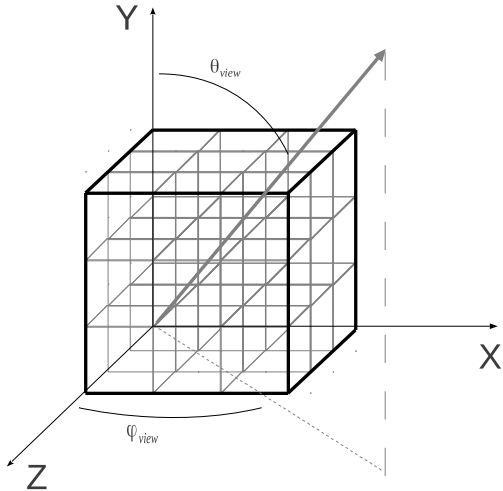}\\    
\end{center}
  \caption{Volume/Jet geometry illustrating how a block of cells can be used to 
construct a larger volume. $\theta_{view}$, the viewing angle with respect to the 
volume/jet is incorporated into Doppler boosting calculations. } 
\label{fig:volGeo}
\end{figure}

In our model the basic volume structure is a cubic cell. This allows
the model to be modular and build an arbitrarily large volume with any
desired anisotropies. Each cell contains a magnetic field plus
electron and photon distributions. The magnetic field can have an
arbitrary orientation and strength in each cell. This means that the overall
volume can be modelled to contain a completely uniform,
partially anisotropic, or pseudo-random \textbf{B}-field. 
The purpose of the present work is to isolate the effects of
the degree of order and orientation of the magnetic field on the
emerging synchrotron emission. Therefore, we choose the simplest
conceivable approach concerning the electron distribution, and do
not take electron cooling into account. This means, we only focus 
on static electron distributions which do not evolve due to 
energy losses. In future work we aim to include self-consistent 
cooling effects which would also allow us to probe how the pitch-angle 
dependence of the synchrotron cooling would give rise to different 
electron distributions in different cells, depending on the magnetic 
field set-up.

The directional information in the electron and photon distributions and the \bfield\ 
is with respect to the cell. In the case of electrons the distribution is a function 
of energy and two angles with respect to the cell (see figure \ref{fig:eleB}). This 
gives us the ability to create anisotropies in the electron distributions as well by 
either having preferential direction for the electrons or by setting up the electron 
distributions differently in various cells. For both
electrons and photons, the distributions' energy grids (Lorentz factor
$\gamma$ for electrons and frequency $\nu$ for the photons) are calculated using
logarithmic binning. Therefore each distribution is modelled using a 3 dimensional 
array with the dimensions of $[\mathrm{bins}_{\gamma/\nu}\times \mathrm{bins}_{\theta}\times 
\mathrm{bins}_{\phi}]$. Figures \ref{fig:eleB} (right) illustrates how various angles 
with respect to the cell are defined. Angles $\theta_{cell}$ and $\phi_{cell}$ run from 
0 to $\pi$ and 0 to 2$\pi$ respectively. In figure \ref{fig:volGeo} we can 
see how the overall volume can be constructed from individual cells. For a given 
viewing angle, the emission from the visible outer layer of cells is combined to 
produce an overall spectrum from an effectively larger volume.

The simulation currently transfers, from one cell to
another, only the photons. In order to achieve this, we need to calculate which of 
the six cubic faces a given photon direction will intersect. 
To calculate 
this, we assume that all the photons are produced in the center of the
cell, and then trace photon paths in any given direction towards the nearest boundary. 
Even though our simulation considers a static situation, the transfer and radiative
feedback between different cells requires an inherent time-dependence in the code. 
The time step for our radiation transfer approach is the light crossing time across
a single cell, which is equal to the time it takes for the photons to travel from 
one cell to another. At the end of each time step and depending 
on the physical processes being modelled, the photon distribution is modified and 
passed to the appropriate neighbour. When being passed to a neighbour the entire 
photon distribution is passed. Therefore at the end of a time-step each cell's 
(intrinsic) photon distribution is emptied into six neighbouring cells, unless 
it is a boundary cell. The six incoming (transiting) photon distributions are stored 
until the start of the following time-step when they are combined to form a single 
intrinsic photon distribution again. The physical processes are then carried out on 
this single photon distribution. Synchrotron radiation is calculated first and the
photons added to the intrinsic distribution. Compton scattering is carried out after 
the synchrotron radiation. At this point we reach the end of a time-step and the 
process of transferring photon distributions to neighbouring cells begins again. 
The observed photon distribution originates from the boundary cells. The 
photon distributions emerging from visible faces of the boundary cells are combined 
to create a single observed photon distribution. This process of combining the photon 
distributions from the boundary cells in effect treats the whole multi-cell structure 
like a single cubic/cuboid structure.

\subsection[]{Synchrotron radiation}
\label{sec:synch}
Here we highlight the key points of the synchrotron radiation
treatment that we follow. A more in-depth analysis and details can
be found in \cite{longair94}.

The synchrotron emissivity per electron, $P_{\nu}$, is given by:
\begin{eqnarray}
\label{eqn:synchPower}
  P_{\nu}=\frac{\sqrt{3}e^3B\sin\alpha}{
    m_ec^2} F(x) ~~,
\end{eqnarray}
where F(x) is given by:
\begin{eqnarray}
\label{eqn:bessel}
F(x) = x  \int_x^{\infty}{K_{5/3}(z)\mathrm{d}z} ~~.
\end{eqnarray}
$x=\nu/\nu_c$, where $\nu_c$ is the critical frequency given by
$3\gamma^2 e B \mathrm{sin}\alpha/4\pi m_e$c. $\alpha$, the pitch angle, is calculated 
using spherical trigonometry: 
\begin{eqnarray}
  \cos(\alpha) &=& \cos(\theta_e)\cos(\theta_B)\ \nonumber \\
  &+& \sin(\theta_e)\sin(\theta_B)\cos(\phi_B-\phi_e) ~~ .
\end{eqnarray}
The synchrotron emission coefficient is given by:
\begin{eqnarray}
\label{eqn:synchEm}
  j_{\nu}(\Omega_{sy})&=&\oint_{4\pi}\mathrm{d}\Omega_e
  \int_1^{\infty}\mathrm{d}\gamma n(\gamma,\Omega_e) \nonumber \\
  &\times& P_\nu~ \delta(\Omega_{sy}-\Omega_{e})~~.
\end{eqnarray}
 The numerical Bessel function integration in equation \ref{eqn:bessel} can be time consuming. 
However, some
fast routines to perform this integration are given by
\cite{umstaetter128501} which we modified for our precision and
computer language.

In a full treatment of the synchrotron radiation the
emitted photons are distributed within a solid angle ($\Omega \sim 1/\gamma^2$) about the pitch
angle $\alpha$. However, for our purposes we assume the emitted photons travel in the same 
direction as the emitting
electrons. 

A detailed calculation of synchrotron self-absorption can be found
\cite{longair94}. The absorption coefficient when recast in terms of electron Lorentz factors,
$\gamma$, instead of $E$, can be written as:
\begin{eqnarray}
\label{eqn:synchAb}
  \chi_{\nu}= -\frac{1}{8\pi
    m\nu^2}\int_1^{\infty}P_{\nu}\frac{\mathrm{d}}{\mathrm{d}\gamma}\left(
  \frac{N(\gamma)}{\gamma^2}\right)\gamma^2\mathrm{d}\gamma ~~.
\end{eqnarray}
The photons produced via synchrotron radiation are added to the intrinsic photon 
distribution of the cell. The photons received from neighbouring cells are added to the 
intrinsic photon distribution prior to calculating the synchrotron spectrum. Therefore 
the photons passing through any cell are also synchrotron self-absorbed. 
The emission 
and absorption coefficients are used to calculate the total spectrum,
\begin{eqnarray}
\label{eqn:synchSp}
I_{\nu}=\frac{j_{\nu}}{4\pi\chi_{\nu}}(1-e^{-\tau_{\nu}}) ~~,
\end{eqnarray}
where $\tau_{\nu} = \chi l$ is the optical depth and \textit{l} is the size of the 
emission zone/cell. 

\subsection[]{Compton scattering}
\label{sec:comp}
\begin{figure}[h]
\begin{center}
  \includegraphics[width=0.4\textwidth]{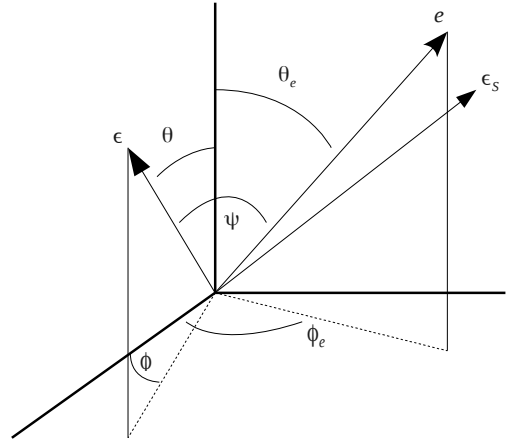}
\end{center}
  \caption{Various angles involved in the Compton
    scattering of photons with energy $\epsilon$ to $\epsilon_s$
    off electrons $e$}
\label{fig:compAng}  
\end{figure}
In the limit $\gamma>>1$, and in the electron rest frame, the incident photon 
travels in nearly the opposite direction to the electron. This is due to photon 
aberration:
\begin{eqnarray}
\label{eqn:aberr}
\cos \theta ' = \frac{\cos \psi - \beta_e}{1-\beta_e\cos\psi} ~~ .
\end{eqnarray}
When $\beta_e\rightarrow 1$, $\cos\theta '\rightarrow -1$, we are in the 
head-on approximation regime, which we employ to greatly simplify Compton
cross section calculations. That is, we can assume that the scattered
photon solid angle, $\Omega_s$, is well approximated by the electron
solid angle $\Omega_e$. 
When the differential Compton (Klein-Nishina) cross section is 
integrated over $\Omega_s$, we get \citep{dermer09}:
\begin{eqnarray}
\label{eq:compCS}
\frac{\mathrm{d}\sigma_C}{\mathrm{d\epsilon_s}}\simeq\frac{\pi
  r_e^2}{\gamma \epsilon '}\Xi_C \mathrm{H} \left(\epsilon_s;\frac{\epsilon
  '}{2\gamma},\frac{2\gamma\epsilon '}{1+2\epsilon '} \right) ~~,
\end{eqnarray}
where H is a Heaviside function, $\epsilon ' =
\gamma\epsilon(1-\beta_e\cos\psi)$, and the Compton kernel is given by:
\begin{eqnarray}
\label{eqn:compker}
\Xi_C\equiv y + y^{-1}-\frac{2\epsilon_s}{\gamma\epsilon '
  y}+\left(\frac{\epsilon_s}{\gamma \epsilon 'y}\right)^2 ~~, 
\end{eqnarray}
and $y = 1 - (\epsilon_s/\gamma)$. The Compton cross section can
then be used in the emission coefficient formula to
obtain the Comptonized spectrum. The head-on approximation simplifies the 
emission coefficient calculation by eliminating two integrals from the 
Compton emissivity treatment without the approximation. The following 
relation can be used to obtain the number of interacting photons: 
\begin{eqnarray}
  n_{ph}(\epsilon)=\frac{1}{hc}\oint\mathrm{d} 
\Omega\int_{\epsilon_{1}}^{\epsilon_{2}}\mathrm{I}_{\nu}(\epsilon,\Omega)
\frac{\mathrm{d}\epsilon}{\epsilon}
\end{eqnarray}
The interacting photons are a combination of photons originating from synchrotron 
radiation and the photons received from neighbouring cells. At the start of a 
time-step, the photon distributions received from the neighbouring cells are 
combined, while preserving the direction information, to form the intrinsic 
photon distribution. Synchrotron photons are also added to the intrinsic photon 
distribution. 
The total photon distribution is then used in the Compton emissivity 
relation to obtain the Compton spectrum in the head-on approximation, given by:

\begin{eqnarray}
  j_C^{head-on}(\epsilon_s,\Omega_s) &=& \frac{m_ec^2}{h}\epsilon_s
  \int_1^{\infty}\mathrm{d}\gamma n_e(\gamma, \Omega_e=\Omega_s) \nonumber \\ 
  &\times& \int_{-1}^{1}\mathrm{d}\mu \int_{0}^{2\pi}\mathrm{d}\theta \int_0^{\infty}\frac{\mathrm{d}\epsilon}{\epsilon} I_{\mathrm{\epsilon}}(\epsilon,\Omega)\nonumber \\
  &\times& (1-\beta_e \cos\psi) \frac{\mathrm{d}\sigma_C}{\mathrm{d}\epsilon_s} \ .
\end{eqnarray}

After the intrinsic photon distribution has been Compton scattered it is 
redistributed based on change in energy and direction. The redistributed 
photon distribution is then used to work out which neighbouring cells receive 
which proportion of the distribution.

\subsection[]{Overall spectrum}
The overall spectrum is obtained by combining the synchrotron and
Compton spectra. As it stands in our model, synchrotron emission is the only 
source of photons which are then Compton scattered by the
same population of electrons. The resulting photon spectrum is given as a 
function of two cell angles $\theta_{cell}$ and $\phi_{cell}$ (see figure 
\ref{fig:eleB}). Although not included in the present
study, it is straight forward to include external Compton effects by
adding the external photon field to the photon distributions.  

Once we have a spectrum, $I_{\nu}$, the flux can be calculated using:
\begin{eqnarray}
\label{eqn:flux}
F_{\nu_{obs}} = \delta_{\mp}^3\frac{A}{4\pi^2d_L^2}I_{\nu_{emit}} ~~,
\end{eqnarray}
for an emission zone with an area $A$ and luminosity distance $d_L$. 
For a viewing angle $\theta_j$ the Doppler factor $\delta_{\mp}$
is given by
\begin{eqnarray}
  \delta_{\mp}=\left[\Gamma(1\mp\beta(\textrm{cos}\theta_j)\right]^{-1} \ .
\label{eqn:doppler}
\end{eqnarray}
The `$\mp$' corresponds to either an approaching or a receding
component of the jet. In the case of blazars the observed emission is strongly 
dominated by the approaching jet, boosted with the Doppler factor $\delta_-$. 
Also, any given frequency, $\nu$, in the emission region rest frame will
be shifted by a factor of:
\begin{eqnarray}
\label{eqn:rshift}
\nu_{obs} = \nu_{emit}\frac{\delta_-}{1+z}
\end{eqnarray}
We follow the \cite{wright06} formulation to calculate the luminosity
distance based on the redshift of an object. Photons emitted at an
angle $\theta_{ph} '$ in the cell rest frame will appear at an angle
$\theta_{ph}$ due to angle aberration, which can be expressed as follows:
\begin{eqnarray}
  \label{eq:angaber}
  \cos\theta_{ph} = \frac{\cos\theta_{ph}'+\beta}{1+\beta\cos\theta_{ph}'} ~~.
\end{eqnarray}

\section[]{Results}
\label{sec:res}
\begin{figure}[h]
  \includegraphics[width=0.45\textwidth]{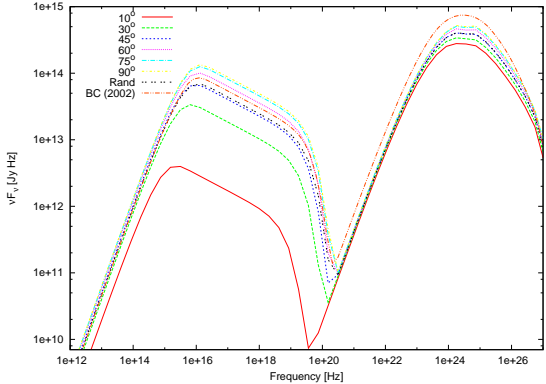}
  \caption{Synchrotron self-Compton spectra at a viewing angle of $0.014^o$ 
($0.57^o$ in jet frame). The spectra shown results of different B-field 
configurations, including when it is randomly oriented in different cells. 
The plot shows $\theta_B$ values ($\phi_{B} = 90^o$, unless randomly oriented).
The red dot-dot-dashed curve shows the corresponding calculation based 
on the angle averaged emissivity and spherical geometry, using the code of 
\cite{boettcher02}.
See figure \ref{fig:eleB} for details on various angles.}
\label{fig:B_angles_ssc}  
\end{figure}
\begin{figure}[h]
  \includegraphics[width=0.45\textwidth]{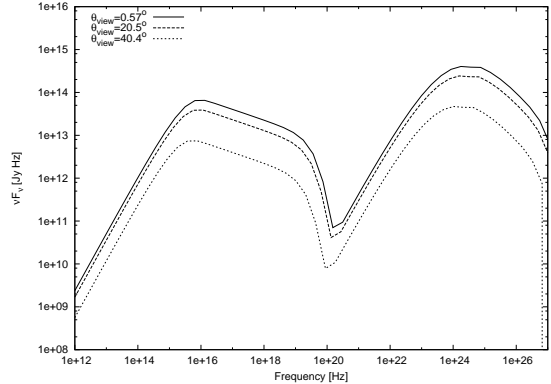}
  \caption{Synchrotron self-Compton spectra at a various viewing angles (jet 
frame values). The \bfield\ orientation: $\theta_B = 45^o$, $\phi_{B} = 90^o$. 
See figure \ref{fig:eleB} for details on various angles.}
\label{fig:view_angles_ssc}  
\end{figure}

In order to study the effects of the \textbf{B}-field orientation on the 
synchrotron and synchrotron self-Compton spectra, we set up two
scenarios. In the first set up the magnetic field is uni-directional
in all the cells (27 in total) and in the second scenario the magnetic
field is randomly oriented in each of the cells. These set-ups are likely to
be the two extreme scenarios for a jet. Evidence points to
the \textbf{B}-field being semi-ordered in AGN jets; for example, helical 
\citep{attridge99,lyutikov05,marscher08}. In all the presented cases, the 
electron distribution is a power-law and distributed uniformly over the angles 
$\theta$ and $\phi$.

Figure \ref{fig:B_angles_ssc} shows synchrotron self-Compton spectra with 
different \bfield\ configurations. Various simulation parameters can be found 
in table \ref{tab:figparams}. The spectra are for a fixed viewing angle. 
Changing the \bfield\ orientation has a significant impact on the observed 
spectrum. There is a large difference in flux values depending on the \bfield\ 
orientation with respect to the observing direction. The minimum flux levels are 
observed at an orientation along the viewing angle while the maximum flux levels 
are observed when the \bfield\ is perpendicular to the line of sight. The figure 
also demonstrates the fact that only the synchrotron spectrum component is heavily 
affected by the \bfield\ orientation. The Compton scattered component of the spectrum 
is almost independent of the \bfield\ orientation. The main reason for this is that 
the photon distribution anisotropies introduced by the \bfield\ orientation are lost
when scattering off an isotropic distribution of electrons. The small variations that 
remain are due to the $1-\beta\cos\psi$ factor present in Compton emissivity calculations. 
The line of sight and photon anisotropies therefore affect the extent to which the 
Compton spectrum is boosted. Additional anisotropies are introduced by the discretization 
of the photon and the electron angular distributions. 

Figure \ref{fig:B_angles_ssc} also shows 
a comparison with another SSC calculation \citep{boettcher02} which assumes a randomly oriented 
magnetic field. This simulation is set up with identical parameters to the ones outlined in table 
\ref{tab:figparams}, except it uses a spherical volume instead of a cube; the total volumes are 
identical, therefore the sphere has a radius of $1.86 \times 10^{14} $ cm. We can see that the
synchrotron components are in good agreement. However, the Compton component in the 
\cite{boettcher02} calculation is much higher. The inverse Compton to synchrotron 
ratio $F_{IC}/F_{syn}$ differs by a factor $\sim$0.68 between the two calculations. 
This is most likely due to differing geometries. For a sphere, the average photon 
escape time is ${3R/4c}$, whereas in our cubic set up, due to the way radiation transport
between cells is treated (see Section \ref{sec:volume}), a photon takes, on average, 
$2d_{cell}/c$ to escape the region ($d_{cell}$ is the width of an individual cell).  
Since the flux ratio $F_{SSC}/F_{syn} = u_{\rm syn}/u_B$ and the volume-averaged 
radiation energy density $u_{\rm syn}$ is proportional to the photon escape time 
scale, the longer photon escape time scale in the spherical geometry results in a 
larger SSC flux.

Figures \ref{fig:view_angles_ssc} shows the effects of the viewing angle on the spectrum. 
It shows SEDs for a fixed azimuthal angle, but different viewing angles with a single 
\bfield\ configuration. As before, there are normalization differences between the SEDs 
when comparing uniform and randomly oriented \bfield, but the main factor in this case 
is the variation in the Doppler boosting due different viewing angles. 
\begin{table}[h]
\begin{center}
\scalebox{1.0}{
\begin{tabular}{ll}
\hline
Parameter & fig \ref{fig:B_angles_ssc} \& \ref{fig:view_angles_ssc}\\
\hline
Cells & 27\\
Cell size  & $1\times 10^{14}$ cm\\
z & 0.031 \\
$\Gamma$ & 20.0 \\
$u_e$ & $9$ ergs cm$^{-3}$\\
p & 3.5 \\
$\gamma_{min}$ & $5\times 10^{3}$\\
$\gamma_{max}$ & $5\times 10^{5}$ \\
B & $0.9 ~\mathrm{G}$ \\ 
\hline
\end{tabular}
}
\caption{The simulation parameters for the results show in figures \ref{fig:B_angles_ssc} 
and \ref{fig:view_angles_ssc}. Key to symbols: z = redshift,  $\Gamma$ = Jet Lorentz 
factor, $u_e$ = electrons energy density, $p$ electrons distribution power law index, 
$\gamma_{min,max}$ = electrons distribution minimum and maximum energy, $B$ = \bfield\ 
strength}
\label{tab:figparams}
\end{center}
\end{table}

\subsection{Markarian 421}
\begin{figure}[h]
  \includegraphics[width=0.45\textwidth]{figure8.epsi}
  \caption{Synchrotron self-Compton fit for Markarian 421. The spectral fits are at 
various \bfield\ orientations. The data are from \textit{XMM-Newton} OM, \textit{XMM-Newton} 
EPIC, and VERITAS. See table \ref{tab:mrkparams} for the fit parameters.}
\label{fig:mrk421_0402}  
\end{figure}
\begin{figure}[h]
  \includegraphics[width=0.45\textwidth]{figure9.epsi}
  \caption{Synchrotron self-Compton fit for Markarian 421. The spectral fits are at 
various \bfield\ orientations. The data are from \textit{XMM-Newton} OM, \textit{XMM-Newton} 
EPIC, and VERITAS. See table \ref{tab:mrkparams} for the fit parameters.}
\label{fig:mrk421_0802}  
\end{figure}
\begin{figure}[h]
  \includegraphics[width=0.45\textwidth]{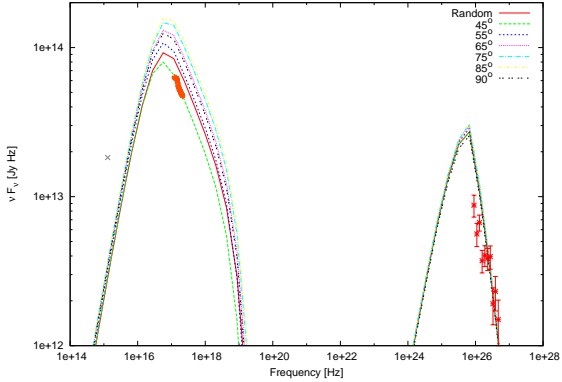}
  \caption{Synchrotron self-Compton fit for Markarian 421. The spectral fits are at various 
\bfield\ orientations. The data are from \textit{XMM-Newton} OM, \textit{XMM-Newton} EPIC, 
and VERITAS. See table \ref{tab:mrkparams} for the fit parameters.}
\label{fig:mrk421_1802}  
\end{figure}

Markarian 421 was the first extragalactic source to be detected in TeV energies, hence 
making it an extensively studied source. We present spectral fits to \textit{XMM-Newton} 
and \textit{VERITAS} data presented in \cite{acciari09}. We refer the reader to the above 
paper for the details on data reduction. 

Figures \ref{fig:mrk421_0402}, \ref{fig:mrk421_0802}, and \ref{fig:mrk421_1802} show 
fits to Mrk 421 data using our code. Our aim in the present paper is to explore the 
impact of the \bfield\ orientation on the fit parameters, especially the \bfield\ 
estimates. In figure \ref{fig:mrk421_0402} we can see that the synchrotron component 
is best fit for a particular \bfield\ orientation (see table \ref{tab:mrkparams} for 
fit parameters). The gamma-ray data, however, is fit well with all the orientations.  
This is due the fact that the self-Compton spectrum is not affected much by the 
\bfield\ orientation (see the discussion in the previous section). There is a 
significant difference in the synchrotron peak when comparing various \bfield\ 
orientations. The spectra in figure \ref{fig:mrk421_0402} show that at a good fit 
is achieved when the \bfield\ is oriented at $85^o$ with a strength of $0.18$~G. 
However, the fit shown is not unique. We can see in figures \ref{fig:mrk421_0802} 
and \ref{fig:mrk421_1802} that for identical parameters, except the \bfield\ strength, 
the best-fit \bfield\ orientation is very different. In one case a pseudo-random \bfield\ 
provides the best fit 
with $B = 0.22$~G, while for $B = 0.25$~G, 
a \bfield\ orientation of $45^o$ provides the best fit. Therefore 
good fits can be achieved with different \bfield\ strengths at different orientations. 
The main point here is the fact that it is possible to over- or underestimate the 
\bfield\ strength when assuming it to be randomly oriented. In the cases presented here, 
the \bfield\ strength ranges from 0.18 G to 0.25 G for very similar fits to the data, 
but with different magnetic-field orientations. 
Therefore it is possible to 
overestimate the magnetic field strength by at least $30\%$ if a particular \bfield\ 
orientation, whether uniform or tangled, is assumed. If the \bfield\ were pointed 
closely aligned with the line of sight, much higher \bfield\ values will be necessary 
to obtain similar fits (see discussion in section \ref{subsec:bestimate}). 
We also note that the bulk Lorentz factor used in the fits are lower than the 
values obtained by some authors for fitting Mrk 421 data \citep[e.g. see][]{aleksic11}. 
The value of the bulk Lorentz factor values used in our fits is likely to be on the 
lower end of the limits imposed by pair opacity arguments \citep{celotti98}. However,
the main point of this paper is not the determination of actual best-fit values for
Mrk~421 (which would not be realistic due to our neglect of cooling effects anyway),
but to demonstrate the orientation-dependent magnetic-field degeneracy in the course 
of blazar SED fitting. 

\begin{table}[h]
\begin{center}
\scalebox{0.7}{
\begin{tabular}{llll}
\hline
Parameter & fig \ref{fig:mrk421_0402} & fig \ref{fig:mrk421_0802} & fig \ref{fig:mrk421_1802}\\
\hline
Cells & 27 & 27 & 27\\
Cell size  & $3\times 10^{15}$ cm & $3\times 10^{15}$ cm & $3\times 10^{15}$ cm\\
z & 0.031 & 0.031 & 0.031\\
$\Gamma$ & 10.0 & 10.0 & 10.0\\
$u_e$ & $0.01$ ergs cm$^{-3}$ & $0.01$ ergs cm$^{-3}$ & $0.01$ ergs cm$^{-3}$\\
p & 4.1 & 4.1 & 4.1\\
$\gamma_{min}$ & $3.7\times 10^{4}$ & $3.7\times 10^{4}$ & $3.7\times 10^{4}$\\
$\gamma_{max}$& $5\times 10^{5}$ & $5\times 10^{5}$ & $5\times 10^{5}$ \\
B & $0.18 ~\mathrm{G}$& $0.22 ~\mathrm{G}$ & $0.25 ~\mathrm{G}$ \\ 
\hline
\end{tabular}}
\caption{Fit parameters for the Markarian 421 data shown in figures \ref{fig:mrk421_0402}
 and \ref{fig:mrk421_0802}. Key to symbols: z = redshift,  $\Gamma$ = Jet Lorentz factor, 
$u_e$ = electrons energy density, $p$ electrons distribution power law index,
$\gamma_{min,max}$ = electrons distribution minimum and maximum energy, 
$B$ = \bfield\ strength}
\label{tab:mrkparams}
\end{center}
\end{table}

\section{Conclusions}
\label{sec:conc}
In this paper we have presented first results from a new relativistic jet radiation
transfer code that we are currently developing. Here we take the full angular 
dependence into account when modelling synchrotron and synchrotron self-Compton 
processes. We are able to model the \bfield\ at arbitrary orientations and study 
its impact on the resulting spectra.

We have seen that the \bfield\ orientation plays an important role on the normalization 
of the synchrotron spectrum. Using fits to Markarian 421 data, we have shown how the 
\bfield\ orientation can mislead into over/under-estimating its strength. Any future 
work should therefore be mindful of the fact that the underlying assumption about the 
\bfield\ orientation will play a considerable role in the errors associated with the 
magnetic field strength estimates.

\bibliographystyle{apj}

\clearpage

\clearpage




\end{document}